# Evidence of ammonium salts in comet 67P as explanation for the nitrogen depletion in cometary comae

**Authors:** K. Altwegg[1*], H. Balsiger[1], J.-J. Berthelier[2], C. Briois[3], M. Combi[4], H. Cottin[5], J. De Keyser[6], F. Dhooghe[6], B. Fiethe[7], S. A. Fuselier[8,9], T. I. Gombosi[4], N. Hänni[1], M. Rubin[1], M. Schuhmann[1], I. Schroeder[1], T. Sémon[1], S. Wampfler[2]

**Affiliations:**

[1]Physikalisches Institut, University of Bern, Sidlerstr. 5, CH-3012 Bern, Switzerland.

[2]Center for Space and Habitability, University of Bern, Sidlerstr. 5, CH-3012 Bern, Switzerland.

[2]LATMOS/IPSL-CNRS-UPMC-UVSQ, 4 Avenue de Neptune F-94100, Saint-Maur, France.

[3]Laboratoire de Physique et Chimie de l'Environnement et de l'Espace (LPC2E), UMR CNRS 7328 – Université d'Orléans, France

[4]Department of Climate and Space Sciences and Engineering, University of Michigan, 2455 Hayward, Ann Arbor, MI 48109, USA.

[5] LISA, UMR CNRS 7583, Université Paris-Est-Créteil, Université de Paris, Institut Pierre Simon Laplace (IPSL), Créteil, France

[6]Royal Belgian Institute for Space Aeronomy, BIRA-IASB, Ringlaan 3, B-1180 Brussels, Belgium.

[7]Institute of Computer and Network Engineering (IDA), TU Braunschweig, Hans-Sommer-Straße 66, D-38106 Braunschweig, Germany.

[8]Space Science Directorate, Southwest Research Institute, 6220 Culebra Rd., San Antonio, TX 78228, USA.

[9]Department of Physics and Astronomy, The University of Texas at San Antonio, San Antonio, TX, 78249

**Cometary comae are generally depleted in nitrogen. The main carriers for volatile nitrogen in comets are $NH_3$ and HCN. It is known that ammonia readily combines with many acids like e.g. HCN, HNCO, HCOOH, etc. encountered in the interstellar medium as well as in cometary ice to form ammonium salts ($NH_4^+X^-$) at low temperatures. Ammonium salts, which can play a significant role in prebiotic chemistry, are hard to detect in space as they are unstable in the gas phase and their infrared signature is often hidden by thermal radiation or by e.g. OH in minerals. Here we report the presence of all possible sublimation products of five different ammonium salts at comet 67P/Churyumov-Gerasimenko measured by the ROSINA instrument on Rosetta. The relatively high sublimation temperatures of the salts leads to an apparent lack of volatile nitrogen in the coma. This then also explains the observed trend of higher $NH_3/H_2O$ ratios with decreasing perihelion distances in comets.**

Main

Nitrogen in the volatile part of a comet nucleus is predominantly in the form of $NH_3$ and HCN, which are on average (0.80 ± 0.20) % and (0.21 ± 0.02) %, respectively relative to water, e.g. (1). The numbers for HCN are somewhat uncertain as IR observations generally differ from radio observations (2). Apart from these two molecules, nitrogen bearing species have rather low abundances in comets (2). Especially, neutral $N_2$ escaped detection before the Rosetta mission. Already in 1988, after the Giotto flyby at comet 1P/Halley, Geiss (3) recognized that, while carbon and oxygen relative to silicon are close to solar abundance, comet Halley was clearly lacking nitrogen. One explanation for this depletion at that time was the high volatility of $N_2$, which may not have been condensed in the cometary ice or may have been lost in the last 4.6 Gy. For comet 67P / Churyumov-Gerasimenko (67P hereafter), neutral $N_2$ has now been found on the level of (8.9 ± 2.4) × $10^{-4}$ relative to water (~3 % relative to CO) (4). Recently, a high $N_2$/CO ratio of 6% has been reported for comet C/2016 R2 (Pan-STARRS) (5). This shows that $N_2$ is condensed and stored in cometary ice, but the reported abundances are by far not enough to explain the deficiency in nitrogen. N/C atomic ratio in the solar photosphere is about 0.3 ± 0.1 (6). In the refractory phase, comets are also depleted in nitrogen with N/C = 0.05 ± 0.03 in comet 1P (7) and N/C= 0.035 ± 0.011 in comet 67P (8).

While the spread in relative abundances of HCN is quite small among comets, the variation for $NH_3$ seems to be much larger (1). What is quite remarkable is the fact that comets with small perihelion distances seem to have much higher $NH_3/H_2O$ values (1). This suggests that ammonia has a higher sublimation temperature in comets than water, although for pure ice sublimation temperatures are 90 K and 140 K, respectively. In comet D/2012 S1 (ISON) between 1.2 and 0.34 AU, Di Santi et al. (9) found an increase in $NH_3/H_2O$ from < 0.78 % up to (3.5 ± 0.3) % and in addition a distribution of $NH_3$, inconsistent with release from the nucleus only. It indicates that ammonia could be in a different chemical form than pure ammonia ice, probably associated to dust. So far there is no unambiguous explanation for this observation. One possibility would be ammonium salts as source for ammonia because they generally have higher sublimation temperatures than water and would therefore fit the above observations.

Ammonium salts are formed between $NH_3$ (base) and acids transferring $H^+$ from the acid to ammonia. The best known are $NH_4^+ Cl^-$ (ammonium chloride), $NH_4^+CN^-$ (ammonium cyanide), $NH_4^+OCN^-$ (ammonium cyanate), $NH_4^+HCOO^-$ (ammonium formate) or $NH_4^+ CH_3COO^-$ (ammonium acetate). All of the involved acids are part of interstellar and cometary ices (e.g. 10). Theoretical calculations show that the proton transfer has no energy barrier. However, molecules have to diffuse and to be correctly oriented (11) which needs slightly elevated temperatures. Grains in astrochemical environments experience temperatures where such mechanism will be possible. This makes all of those salts likely candidates for interstellar ice and therefore of comets. For a list of experimental evidence of ammonium salt formation at low temperatures and their sublimation see Supplementary Table 1.

In the interstellar medium, signatures of an ice band at 4.62 μm were detected by the infrared space observatory ISO towards different protostellar objects (12). These signatures were often identified as originating from XCN. However, this result was heavily debated, and it was concluded that the signatures are most probably due to $[OCN]^-$, the possible anion of ammonium cyanate (e.g. (13) and references therein). Recently, signatures of ammonium carbonate or ammonium chloride have been found on the surface of Ceres (14). For comet 67P, $NH_4^+$ and COOH have been proposed as part of the broad feature at 3.2 μm seen by the Visible and Infrared Thermal Imaging Spectrometer VIRTIS (15) early in the mission, although the observations at that time were not yet conclusive (16).

It is not easy to detect ammonium salts in comets. OH in minerals and thermal radiation mostly hide their infrared signature in refractories. Looking for the products of sublimation of ammonium salts in the coma is not straight forward either as these salts, upon sublimation, dissociate mostly back into ammonia and the respective acid (Supplementary Table 1). The problem therefore is to distinguish the products of ammonium salts from the pure acids and ammonia, stored in cometary ice.

The ROSINA-DFMS mass spectrometer (17) on the Rosetta orbiter has measured many of the volatiles in the vicinity of comet 67P. Notably, it measured $NH_3$ and HCN, from 3.8 au pre-perihelion to 3.8 au post-perihelion with the perihelion distance at 1.25 au. For a short description on how measurements were performed and about data analysis, see Methods. $NH_3$ followed water quite closely, but not exactly. Figure 1a shows the ratio $NH_3 / H_2O$ over the mission duration with perihelion at day 378. A mean value, thought to be representative of the bulk composition, was derived from the period in May 2015 (around days 290-310). For this period, the comet was at heliocentric distances smaller than 2 AU, the Sun in the southern hemisphere and before the close perihelion period with a lot of short lived outbursts. Measurements during this period yielded relative abundances of $NH_3/H_2O$ = (0.66 ±0.20) % and $HCN/H_2O$ = (0.14 ± 0.04) %, respectively (18), in line with the average comet (1). From figure 1a it is evident that during perihelion the $NH_3/H_2O$ ratio is higher than further from the Sun and that there is quite some scattering in the ratio, reaching values of almost 3 %, uncorrelated with latitude/season (fig. 1b). This scattering could be related to dust in the coma, as it coincides with periods when dust was observed by ROSINA COPS as well as by Giada (19). Gasc et al. (20) looked at the outgassing pattern around the second equinox in March 2016 at 2.7 AU. $NH_3$ showed a similar pattern with latitude as water with a sublimation peak at the subsolar latitude of 0°, steeply decreasing with heliocentric distance, while HCN showed a similar behavior as $CO_2$, being released predominantly from the southern hemisphere. This result is quite surprising as the sublimation temperature of HCN (120K) is closer to water than that of $NH_3$ (90K). It therefore seems that either $NH_3$ is very well embedded in the water ice matrix, much better than HCN, or that it is in a chemical state, which has a higher sublimation temperature than the pure ammonia ice. This finding is compatible with ammonium salts.

Looking very closely at the sublimation process reveals some tracers, which can be used to detect the existence of the salts by mass spectrometry. Generally, simple, direct mass spectrometry cannot distinguish isomers. In the following we therefore use in the text by default the name of the most stable molecule, although we are aware that measured signals could also be from associated isomers. Recently, Hänni et al. (21) studied the sublimation of $NH_4Cl$ and $NH_4COOH$ with high resolution mass spectrometry in the lab. The sublimation pattern for ammonium chloride is relatively simple, yielding mostly $NH_3$ and HCl. However, a clear signature of $NH_4^+$ could be detected as well as $NH_2Cl$ (chloroamine). For ammonium formate the signature is much more complex. Apart from $NH_3$ and formic acid, formamide, amines, HNCO and/or HOCN and HNC and/or HCN are formed as well. By measuring the products with a mass spectrometer, there are in addition fragments of all these species due to electron impact ionization in the ion source of the mass spectrometer. Fragmentation patterns for the ammonium salts considered here can be found in Methods.

Sublimation temperatures vary widely between the salts (Supplementary Table 1). As sublimation temperatures for salts are generally higher than for water, it is not expected that substantial amounts of salt sublimate directly from the comet nucleus, but rather from grains in the coma. A dust event towards the end of the Rosetta mission enabled ROSINA to detect many tracers for ammonium salts despite their low densities in the coma.

**Dust event:** Towards the end of the Rosetta mission the spacecraft performed the closest ever orbits around the nucleus. They consisted of ellipses, which were kept constant in size but where the pericenter was slowly lowered, reaching 1.9 km above surface on September 5. The period of the ellipses was exactly three days. At that time, the spin period of the comet was very close to 12 h. That meant that, for all ellipses, the spacecraft had the closest encounter over the same sub-spacecraft latitude and longitude of the comet. At that time, $CO_2$ was clearly dominant, still sublimating preferentially from the southern (winter) hemisphere, while water followed the subsolar latitude, which at that time was at 18° north. The pattern from ROSINA-COPS is very repetitive (Extended Data Figure 2) with diurnal and latitudinal variations, with a slow increase of the peak intensity due to smaller pericenter distances, showing that the outgassing of the comet at that time was very stable, at least up to Sept. 5, 17h. On September 5, 2016, 18h (day 766) at 3.7 AU from the Sun the spacecraft including the ROSINA sensors were hit by dust from the comet. Details on the dust impact are given in the Methods.

The next full mass spectrum started at 20:10 UTC. Shortly after 20 h the COPS signal returned to nominal. Overlaying DFMS mass spectra for *m/z* 17 and *m/z* 36 (fig. 2) shows that spectra taken on August 30, 18:xx h (the exact time is given in the plots and differs from mass to mass because of sequential measurements), Sept 2, 18:xx h and Sept 5, 17:xx h are almost identical as in all cases DFMS was in the same measurement mode, the spacecraft was over the same latitude and longitude of the comet and at a very similar distance from the nucleus. However, 1h later, on September 5, 18:xx h, intensities were higher by a factor 100. While for *m/z* 17, relative intensities between OH (ionization fragment of $H_2O$) and $NH_3$ remained almost constant, the isotopologues of $CH_4$ disappeared. For *m/z* 36 the ratio between $H_2^{34}S$ and $C_3$ remained constant, while argon disappeared and $H^{35}Cl$, which was not detectable in the undisturbed coma towards end of mission, became the highest peak at 18:18 h on Sept. 5 on this integer mass. While *m/z* 17 was measured before the dust grain hit the DFMS ion source (18:07), *m/z* 36 was measured clearly after the impact (18:18). Shortly after 20h on Sept. 5, OH and $H_2^{34}S$ were back to the same level as before the dust event or slightly below, while $NH_3$ and HCl stayed much higher. Ar, $CH_4$ and $C_3$ remained depleted. From these observations we conclude that the dust hitting the spacecraft was mostly devoid of highly volatiles like Ar or $CH_4$. The dust grain(s) entering DFMS contained abundant $NH_3$ and HCl, which sublimated very slowly over the course of many hours at 273 K, much slower than e.g. water (see fig. 1c). Both species were below the detection limit in less than a day, see also Extended Data Figure 4. Similar observations could be made for many species up to *m/z* 60 (Fig. 3). A full list of observed peak heights for the three periods are given in Supplementary Table 3.

As conditions were much more stable around 20:xx h UTC than during the dust impact itself, we compare relative abundances from 20:xx h with data from 17:xx h, normalizing them to water measured at these times. Figure 4 shows the results. In cyan are species which are possible products of the ammonium salts $NH_4Cl$, $NH_4CN$, $NH_4OCN$, $NH_4HCOO$, and $NH_4CH_3COO$ (see Supplementary Table 2). In orange are species, which are uncorrelated to ammonium salts. Green are sulfur bearing species. Species with a circle were not detected in the undisturbed coma at 17:xx h. The ratios given are therefore lower limits. Species with a dark triangle in the center are products of hydrogenation / protonation.

We find the following tracers for ammonium salts in the mass spectra after the impact:

$NH_4^+Cl^-$: There is a significant enhancement of $H^{35}Cl$ and $H^{37}Cl$. HCl was below the detection limit in the undisturbed coma. There is a clear detection of $NH_4^+$. $NH_2Cl$, a tracer for the salt is not detected, as it is, according to (21), below the detection limit.

$NH_4^+CN^-$: The contribution to CN and HCN from cyanide is too low to be clearly seen. The signal on these two species is already very high in the normal coma, reaching at that time of the mission up to 30% relative to water, sublimating together with $CO_2$ mostly from the cold, southern hemisphere. An additional signal from ammonium cyanide or ammonium formate may therefore not compensate the lower intensity of the ambient coma on *m/z* = 26 and 27 at 20:xx h. However, the detection of cyanamide $NH_2CN$, which is very rare in the undisturbed coma, is most probably a solid tracer for the ammonium cyanide.

$NH_4^+OCN^-$: A small part of HNCO is probably due to ammonium formate, but most of it has to come from cyanate, as the ratio measured for HNCO/HCOOH ≈ 1.4, while the ratio expected from ammonium formate sublimation is less than 0.01 (23). In the spectrum of 18:34 h (Extended Data Figure 1), $NH_4OCN$ and/or $(NH_2)_2CO$ are present, an additional possible tracer of the cyanate.

**$NH_4^+HCOO^-$:** Formic acid is clearly enhanced together with formamide and HNCO, all tracers for the ammonium formate.

**$NH_4^+CH_3COO^-$:** $C_2H_4O_2$ (*m/z* 60) is, within error limits, not enhanced, but also not depleted. There are two competing stable isomers with the same formula, acetic acid and glycol aldehyde, which can be distinguished by their main fragments, namely $CHO_2$ / $C_2H_3O$ and $CH_3O$, respectively (Methods). Looking at the undisturbed coma before impact, there is no detection of $CHO_2$ while $CH_3O$ is more than a factor 10 higher than $C_2H_4O_2$. $CH_3O$ may partly be due to methanol, but the absence of $CHO_2$ clearly supports the notion that most of $C_2H_4O_2$ is glycol aldehyde in the undisturbed coma. After the impact $CHO_2$ is quite abundant and $C_2H_3O$ strongly enhanced. $CHO_2$ is also a fragment of formic acid (table S2). However, from fragmentation of formic acid we expect a ratio of $CHO_2$ / $HCOOH$ = 0.78, while the measured ratio is 2.6 (Supplementary Table 3). From the measured abundances of formic acid, of $C_2H_4O$ and of $CH_2O$ and using the fragmentation pattern we deduce that ~75% of the $C_2H_4O_2$ abundance after the impact is due to ammonium acetate.

**Amines:** Amines and their fragments are surprisingly abundant, probably due to hydrogenation from hydrogen released upon sublimation of the salts, also seen in laboratory experiments (21).

It is clear from figure 4 that species possibly correlated with ammonium salts are highly enhanced relative to water after the dust impact compared to before, while most uncorrelated species are less abundant after the impact than in the normal coma. This apparent depletion is probably due to a loss in sensitivity of DFMS as dust grains were still partly blocking the entrance to the analyzer at 20:xx h. $NH_3$ is enhanced by almost a factor 100 compared to before the impact, bringing it to the level of $H_2O$ (fig. 1c). What is also striking is the amount of hydrogenated / protonated species like protonated methanol and $H_2S$ and amines including their fragments ($CH_mN$, m=2…5, $C_2H_nN$, n=4…7). Together with the fact that all the species and fragments related to salts persist for a long time in the ion source at 273 K makes it highly likely that their parents are indeed ammonium salts.

Using HCl as proxy for the ammonium chloride, HNCO for the cyanate, formic acid for the formate, and one of the major fragments of acetic acid $CHO_2$ for the acetate, the rel. abundances are 5:1:1:1. However, as they have very different sublimation temperatures it may also mean that most of the cyanate and formate have already sublimated two hours after the impact.

In the spectrum taken two hours after the impact heavy organics (*m/z* ≥ 50) and some organo-sulfur species, but not $H_2S$, OCS or $CS_2$ are enhanced as well. This result means that, apart from the ammonium salts, there are most probably other species with low volatility embedded in the dust, which sublimate only slowly at the ion source temperature of 273 K.

Discussion
During a dust impact towards the end of mission, ROSINA-DFMS detected all species/fragments typical of the ammonium salts $NH_4^+Cl^-$, $NH_4^+CN^-$, $NH_4^+OCN^-$, $NH_4^+HCOO^-$, and $NH_4^+CH_3COO^-$ over a period of several hours. From these measurements, we can deduce some relative abundances with large error bars, but it is difficult to infer the overall abundance of ammonium salts in comets, as this was a single incident. However, it is known that comets with small perihelion distances have higher $NH_3/H_2O$ ratios than comets which stay further from the Sun (1). These ratios can reach up to 5% at 0.4 AU (1). For comet Halley at 0.9 AU the ratio was 1.5% (22), while for 67P this ratio at 1.25 AU is ~ 0.7 % and ~ 0.25 % outside of 2 AU (fig. 1). Assuming that comets or at least dust grains in the coma inside of 0.5 AU from the Sun are warm enough to sublimate all salts and that for all other comets part or all of the ammonium is missed by measuring coma abundances, then this suggests that up to 90% of $NH_3$ is in the form of salts. This is supported by the distributed source seen for $NH_3$ for comet D/2012 S1 (ISON), 0.34 AU from the Sun (9). It implies that most of the halides, a small part of HCN and HNC, and larger parts of HNCO, formic and acetic acid observed are also released from the corresponding salts. This result then explains the distributed source of HCl found by de Keyser et al. (23). Ammonium formate may also be the reason for, at least partly, the distributed source of HNC, observed with radioastronomy (e.g. (24); (25); (26); (27)) as HNC is likely a product of ammonium formate. Ammonium cyanide is a source of HCN. But due to its low sublimation temperature of 140 K, similar to water, the scale length of sublimation is too short compared to typical scale lengths observable with remote sensing. However, the ammonium salts can explain some of the increase of HCN vs. $H_2O$ seen in comet D/2012 S1 (ISON) close to the Sun (9).

The sublimation of ammonium salt presents a formation pathway for some of the observed species in cometary coma and interstellar clouds like formamide, acetamide, amines and HNC. Ammonium salts are probably starting molecules for more complex prebiotic molecules like urea or glycine, the latter found in 67P (28). The presence of salts also explains why certain species like the halides or HNCO are hard to detect in interstellar clouds or star forming regions. If we assume that much of the halides are in the form of salt, then their high sublimation temperatures would not allow detection in the gas phase. This could also be an explanation for the relatively low interstellar abundances of HNCO.

The series of compounds detected by ROSINA during the Sept. 5th 2016 dust event strengthens the cometary inventory of molecules with an astrobiology relevance. Indeed, ammonium salts can have different interests in the frame of prebiotic chemistry. The most straightforward being its use in amino acids formation as a source of $NH_3$ in Strecker or Bucherer Bergs reactions (see (29)). They are also involved in the formation of amino acids through amination of ketoacids in presence of both oxidized and reduced iron, in a context of hydrothermal prebiotic chemistry studies (30); (31). It is also noteworthy to highlight that ammonium cyanide ($NH_4CN$) is the key reagent in the formation of the nucleobase adenine, a milestone mechanism in prebiotic chemistry (32).

Recent works have highlighted that ammonium salts reacting with glycol aldehyde can promote the formation of proteinogenic amino acids, while cyanamide reacting with glycol aldehyde results in the formation of natural nucleotide (33); (34). It is of great interest that these two key compounds (ammonium salts & cyanamide) are being detected together in the dust particles of 67P while glycol aldehyde is detected already in the undisturbed coma. Glycol aldehyde was also detected earlier in the coma of comet C/2014 Q2 (Lovejoy) (35).

Assuming that the mean abundance of ammonia in comets is indeed what is observed for comets with small perihelion distances, namely ~4% relative to water (fig 5, upper panel, data from Dello Russo et al. (1) and references therein), would then also explain most of the nitrogen deficiency (fig. 5, lower panel) in comets. This figure is adapted from Rubin et al. (18) (originally Geiss (3)). Taking the values measured by Rubin et al. (18) for the ice and the refractory nitrogen and silicon for 67P by Fray et al. (8) and assuming a 1:1 dust to ice ratio by mass (36), yields a nitrogen depletion, which is even stronger for 67P than for 1P/Halley, in line with the larger perihelion distance of 67P compared to 1P/Halley. Correcting this value in both comets to 4% ammonia relative to water (denoted N*), therefore including the ammonia hidden in the ammonium salts, brings the nitrogen abundance close to solar.

**Correspondence and requests for material to:** Kathrin Altwegg (altwegg@space.unibe.ch)

**Acknowledgements**
ROSINA would not have produced such outstanding results without the work of the many engineers, technicians, and scientists involved in the mission, in the Rosetta spacecraft team, and in the ROSINA instrument team over the last 20 years, whose contributions are gratefully acknowledged. Rosetta is an ESA mission with contributions from its member states and NASA. We acknowledge herewith the work of the whole ESA Rosetta team. Work at the University of Bern was funded by the State of Bern, the Swiss National Science Foundation (SNSF, 200021_165869 and 200020_182418), the Swiss State Secretariat for Education, Research and Innovation (SERI) under contract number 16.0008- 2, and by the European Space Agency's PRODEX Program. SFW acknowledges the financial support of the SNSF Eccellenza Professorial Fellowship PCEFP2_181150. JDK acknowledges support by the Belgian Science Policy Office via PRODEX/ROSINA PEA 90020. SAF acknowledges JPL contract 1496541. Work at UoM was supported by contracts JPL 1266313 and JPL 1266314 from the US Rosetta Project. HC is grateful to M. Powner for fruitful discussions about prebiotic chemistry during the preparation of the manuscript.

**Author contributions**

KA was principal investigator of the ROSINA instrument, analyzed the data and wrote part of the paper. HB, JJB, MC, JdK, BF, SFA, TIG contributed hardware to the instrument. MR, FD, MS, IS, TS operated and calibrated the instrument. NH did laboratory experiments on salts and added to the chemistry. CB, HC and SW contributed the part of the paper about the interstellar and astrobiological consequences. All authors read and commented on the paper.

**Author information**

*Physikalisches Institut, University of Bern, Sidlerstr. 5, CH-3012 Bern, Switzerland*
Kathrin Altwegg, Hans Balsiger, Nora Hänni, Martin Rubin, Markus Schuhmann**,** Isaac Schroeder, Thierry Sémon

*Center for Space and Habitability, University of Bern, Sidlerstr. 5, CH-3012 Bern, Switzerland*
Susanne Wampfler

*LATMOS/IPSL-CNRS-UPMC-UVSQ, 4 Avenue de Neptune F-94100, Saint-Maur, France*
Jean-Jacques Berthelier

*Laboratoire de Physique et Chimie de l'Environnement et de l'Espace (LPC2E), UMR CNRS 7328 – Université d'Orléans, France*
Christelle Briois

*Department of Climate and Space Sciences and Engineering, University of Michigan, 2455 Hayward, Ann Arbor, MI 48109, USA.*
Mike Combi, Tamas I. Gombosi

*LISA, UMR CNRS 7583, Université Paris-Est-Créteil, Université de Paris, Institut Pierre Simon Laplace (IPSL), Créteil, France*
Hervé Cottin

*Royal Belgian Institute for Space Aeronomy, BIRA-IASB, Ringlaan 3, B-1180 Brussels, Belgium.*
Johan De Keyser, Frederik Dhooghe

*Institute of Computer and Network Engineering (IDA), TU Braunschweig, Hans-Sommer-Straße 66, D-38106 Braunschweig, Germany*
Björn Fiethe

*Space Science Directorate, Southwest Research Institute, 6220 Culebra Rd., San Antonio, TX 78228, USA and Department of Physics and Astronomy, The University of Texas at San Antonio, San Antonio, TX, 78249*
Steven A. Fuselier

Figure 1: NH$_3$/H$_2$O abundances from Aug. 1, 2014 to June 30, 2016

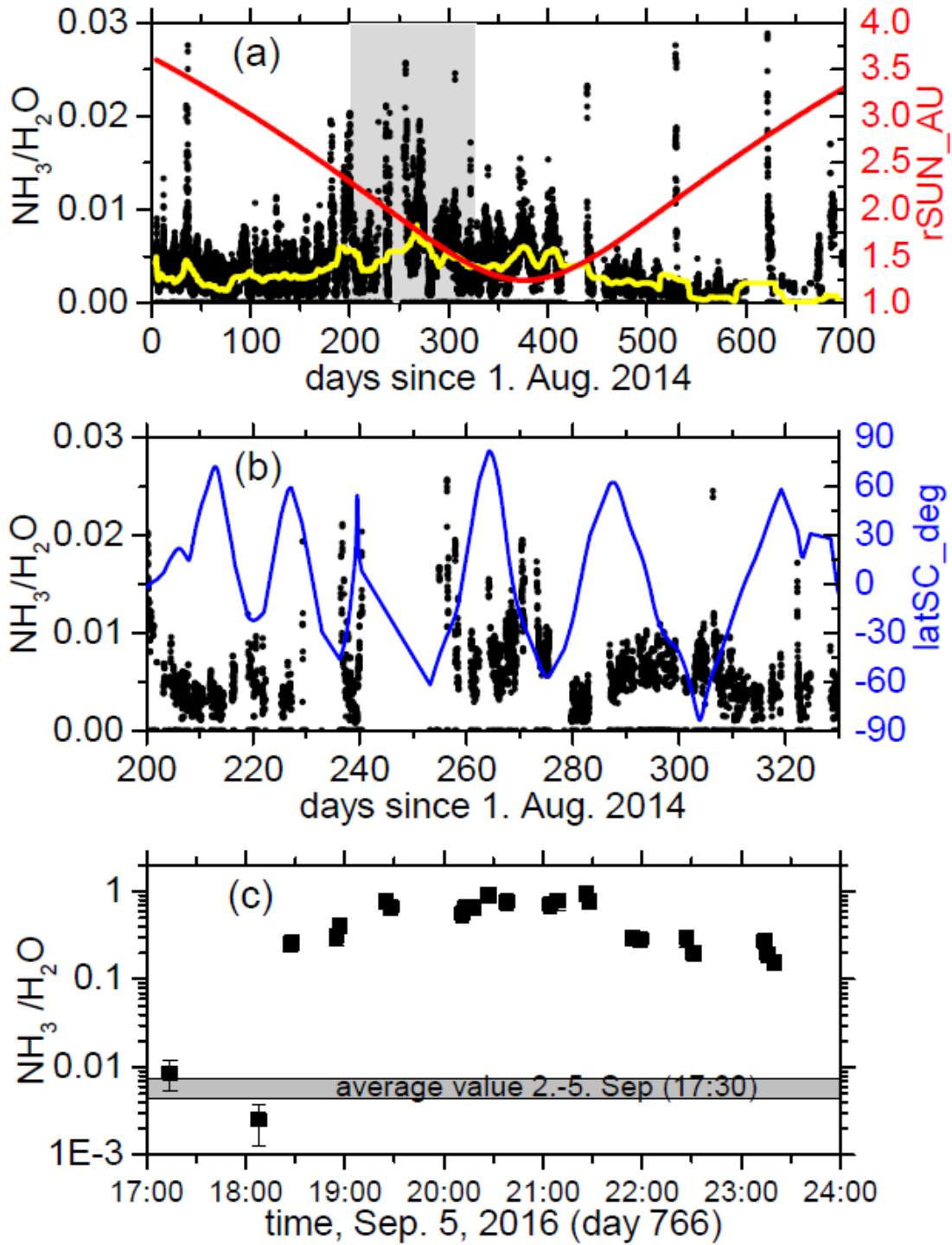

(a) Measured abundance ratio in black. The yellow line shows a moving average, in red: heliocentric distance;
(b) Zoom corresponding to the shaded area in (a), Jan. 28 - Mar 29, 2015. In blue: sub-spacecraft latitude.
(c) Ratio on Sept. 5, 2016 (day 766) and average 2. Sept, 11:00 -5. Sept. 17:30. Error bars: 1-σ statistical errors.

**Figure 2: DFMS mass spectra for *m/z* 17 and 36**

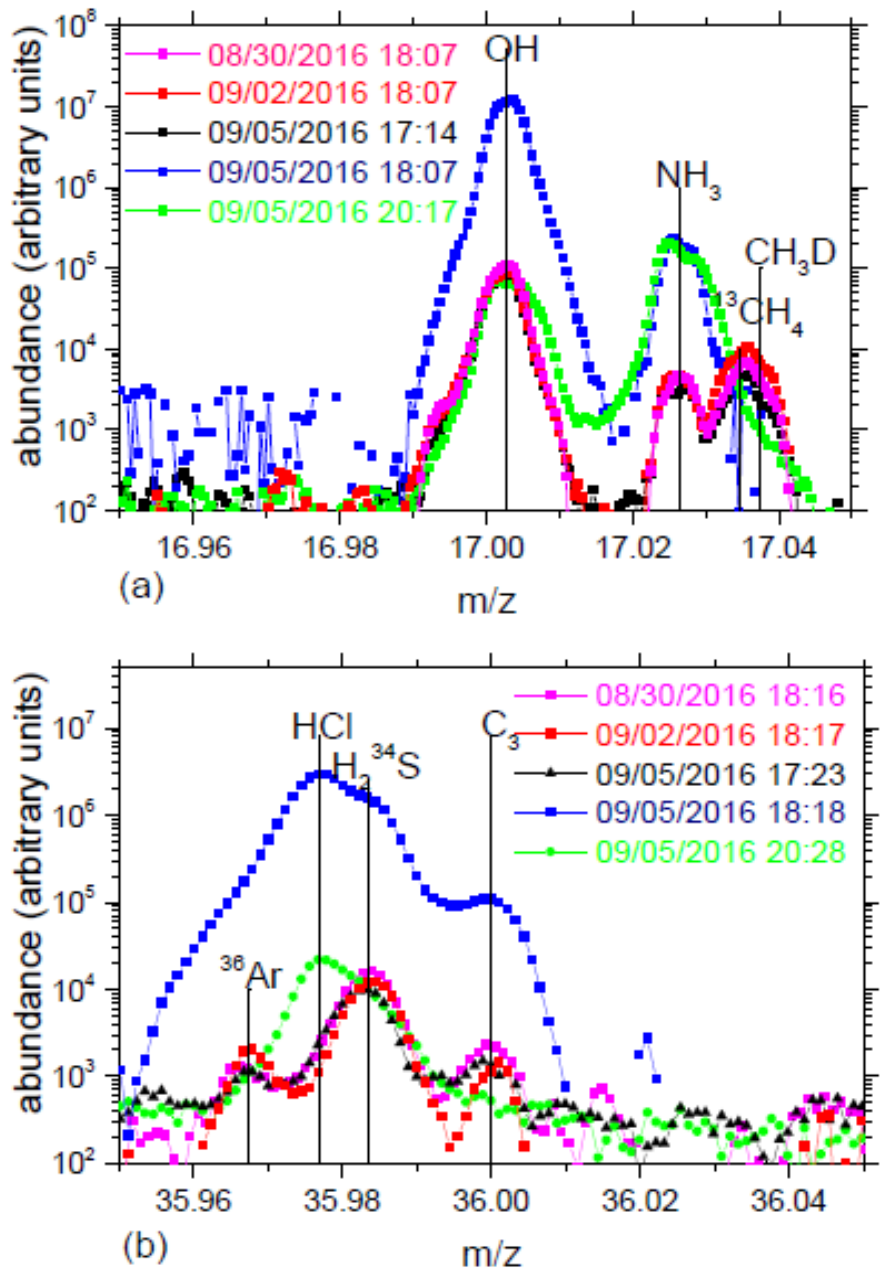

Spectra close to pericenter of end of mission ellipses (Aug.-Sept. 2016) for *m/z* 17 (a) and 36 (b). Exact times in UTC. Orange, violet and black data are three days apart, before dust impact, blue during impact and green after impact. No error bars for clarity reasons. For typical statistical errors see fig. S1.

**Figure 3: DFMS mass spectra before, during and after dust impact**

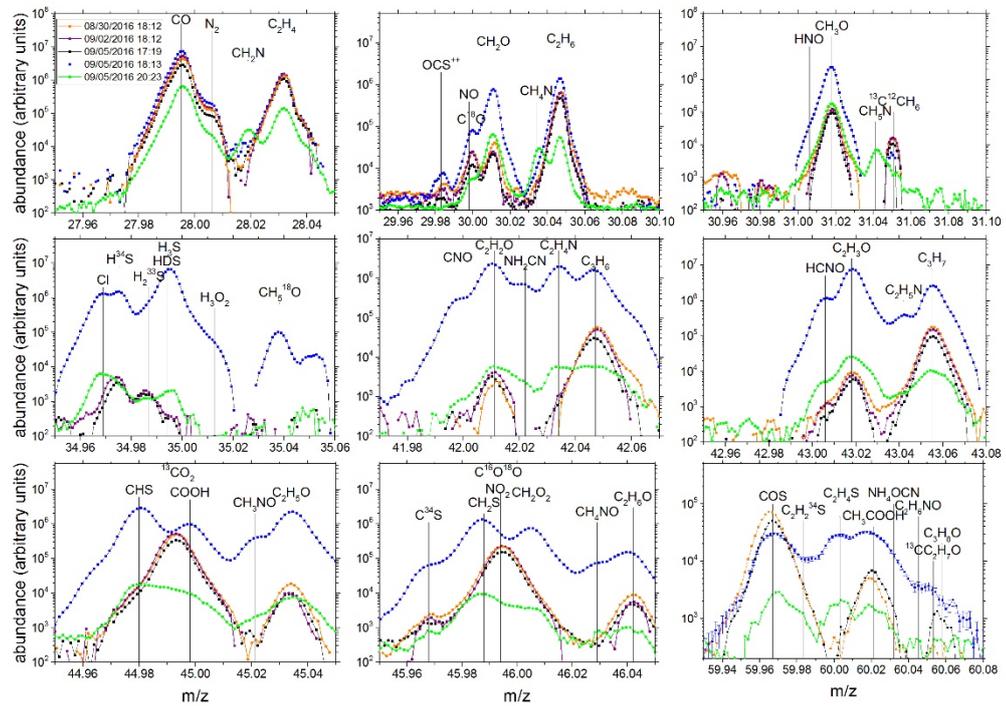

Mass spectra, which exhibit not only a change in intensity, but also in species for Aug. 30, Sept. 2 and the three periods on Sept. 5, 2016. Colors as in fig. 2.

**Fig. 4: Abundance ratios normalized to water**

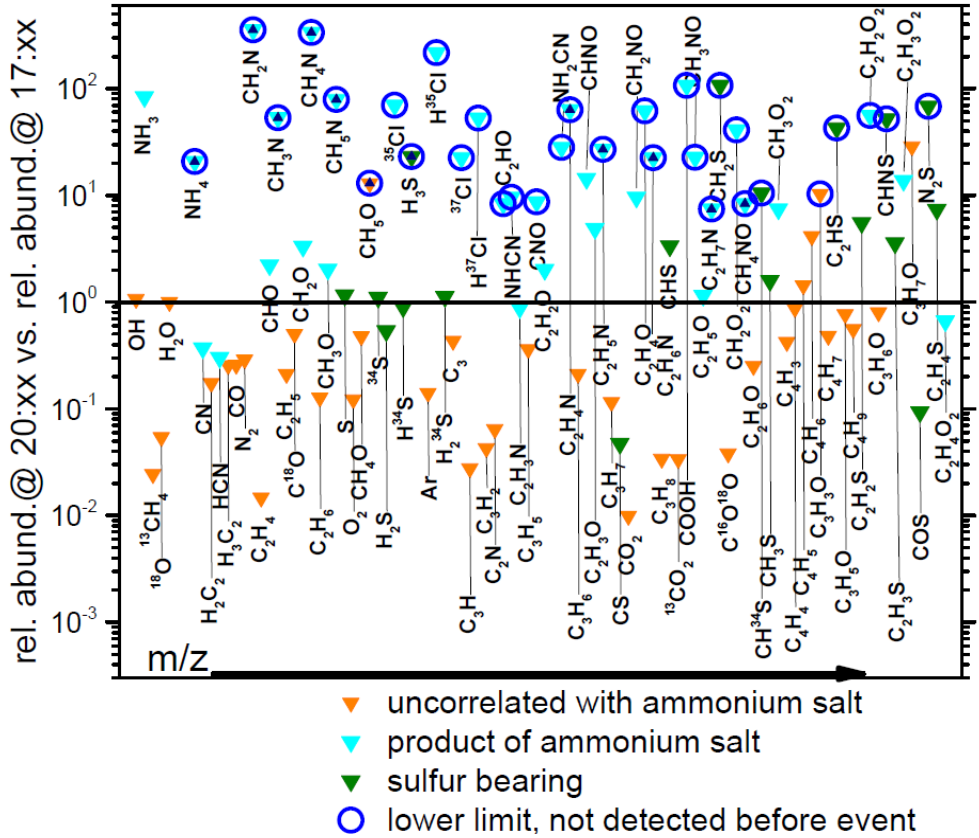

Abundance ratios for the period around 20h compared to 17h UTC on Sept. 5, 2016. For species not detected before impact, upper limits were derived from the noise floor of the detector, which translates into lower limits for the ratios. Sublimation of ammonium salts may produce water, CO and $CO_2$. Their contributions to these species are small as these are the dominant species of the undisturbed coma. We consider them "uncorrelated" to ammonium salt.

**Fig. 5: Relative ammonia and elemental abundances in comets**

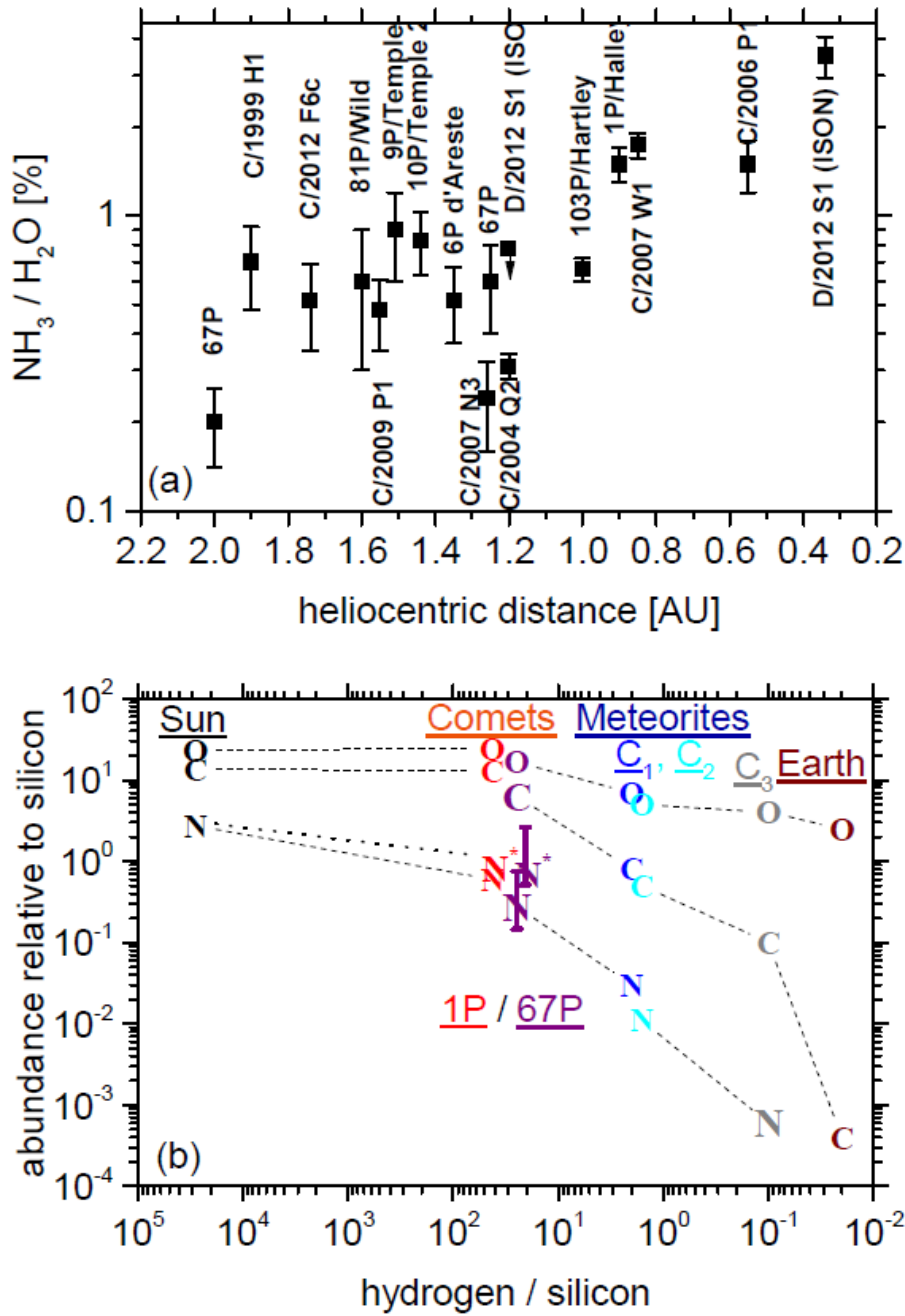

(a) NH$_3$/H$_2$O as a function of heliocentric distance (after Dello Russo et al. (1) with 1-σ statistical errors).
(b) elemental abundances in comets Halley (red) and 67P (purple) for a refractory to ice ratio of 1:1, compared to the Sun, meteorites (C1, C2, and C3, refer to type 1, type 2, and type 3 carbonaceous chondrites) and the Earth (after Rubin et al. (18)). N* denotes the nitrogen abundance if we assume that most of the nitrogen is in ammonium salt and the dust/ice ratio is 1:1 (see text for an explanation).

## Methods

### Measurement modes of ROSINA DFMS

ROSINA DFMS is a classical double focusing magnetic mass spectrometer in Nier-Johnson configuration containing a toroidal shaped electrostatic analyzer and a curved permanent magnet. A full description is given in Balsiger et al. (17). The ions are produced by electron impact ionization from a filament, which is well shielded from the ionization region. Electrons have an energy of 45 eV. Ions are deflected by 7° after passing a relatively broad ion source exit slit in order to prevent dust grains from blocking the narrow entrance slit (14 µm) of the analyzer. The electrostatic analyzer filters the energy (± 1 %), the magnet the momentum. After passing a zoom optics ions hit the Multi-Channel Plate (MCP) detector, which is followed by a linear detector array giving the spatial information with 2 x 512 pixels. The detector is run in analogue mode, which means the gain is adjusted for each measurement within 16 gain steps. The mass resolution is 3000 at the 1% level (~9000 FWHM) of the peak for $m/z$ 28. The high resolution means that below $m/z$ ~50 the width of the detector covers less than one integer mass. The instrument therefore has to step through the masses by adjusting the energy of the ions. It covers a mass range from 12 Da to 180 Da. Integration time per mass is set to 20s. With the overhead (adjustment of voltages, adjustment of the gain) the time needed per mass is ~30s. As the comet outgassing may not be stable over long times, we used in the period considered in this paper four different modes: m502 covering $m/z$ 13 to $m/z$ 50 (~20 min), m562 covering $m/z$ 44 to $m/z$ 100 (~30 min), m564 covering $m/z$ 80 to $m/z$ 140 (~30 min), and m566 covering $m/z$ 135 to $m/z$ 180 (~28 min). Each mode is preceded and followed by a measurement of $m/z$ 18. Due to the overlaps of masses and to the frequent measurements of $m/z$ 18 and using the total density from ROSINA COPS (17) temporal changes in comet outgassing can therefore be monitored. During the period considered in this paper, we used the following sequence of modes: m502, m562, m502, m562, m564, repetitively, which took ~3 h per sequence.

### Analysis of DFMS data

Details on data analysis for DFMS are given in (37) and (38). The peak shape on the detector can be described very well by a double Gaussian, whereby the second Gaussian has a width about three times the width of the first one and the amplitude is < 10% of the peak amplitude. The width is not dependent on the location on the detector as long as it's relatively close to the center. That means if we have several peaks we can use the same width for all of them. Once species have been identified we also know their exact masses. This allows to disentangle even several masses close together with high confidence. An example is given in Extended Data Figure 1 for $m/z$ = 60 where we fit 7 masses, all having the same width. In this case the second Gaussian is mostly unimportant as it affects only the tails of the peaks. But its effect can be seen on the left side of the plot as the tail of COS. The error bars in the plot represent the statistical error from the number of ions. From such fits we get the peak heights and the peak areas. In the present analysis we use peak height in arbitrary units and correct them for the mass (energy) dependent sensitivity of the instrument.

In mass spectrometry, all species registered on the detector are ions, be it primary ions or products from ionization in the instrument. ROSINA was run in neutral mode which has a

+200V potential at the entrance inhibiting primary ions, that is ions already present in the coma, from entering the sensor. $NH_4^+$ is therefore a product, which was created in the ion source itself. Hänni et al. (21) describe possible formation mechanisms of $NH_4^+$ in their paper. In order to make the reader aware that we look at primary neutral species, even if they are being charged in the instrument, we keep $NH_4$ as well as e.g. $H_3S$ without "+" in the paper, both originating as ions from within the sensor to distinguish them from primary ions.

Fragmentation

Ionization in DFMS is done with electron impact at 45 eV. This is lower than in most reference tables, e.g. NIST (39). We expect, however, no big difference in the fragmentation pattern. If at all, the abundance of the parent ion in our case might be enhanced with respect to the daughter fragments. In the case of ammonium salt, we have different species from sublimation before ionization which are then ionized. E.g. for ammonium formate we get the parent molecules $NH_3$, HCN, HNCO, $NH_3CO$, $CO_2$ and HCOOH which are yielding all the fragments in Supplementary Table 2 upon ionization from C to HCOOH. Normally, fragmentation patterns in literature are normalized to the highest fragment which is not necessarily the parent ion. In our case, we normalize it to the acid belonging to the salt.

The fragmentation pattern in low resolution (only integer masses) for ammonium acetate can be found in NIST (39). Ammonium cyanide is not easy to study in the lab due to its poisonous nature. We assume a similar fragmentation pattern as for $NH_4Cl$ and therefore use $NH_2CN$ as tracer. Normally, the ammonium parent salts are not observed. $NH_4OCN$ appears to be an exception. Martinez et al. (40) published a mass spectrum upon irradiation of $NH_3$ + CO ice revealing $NH_4OCN$. However, its detection is ambiguous as it could also be the more stable urea $(NH_2)_2CO$ as ammonium cyanate is a precursor for this molecule (41). We nevertheless adopt this fragmentation pattern also in our case for $NH_4OCN$. A list of species / fragments from the sublimation of ammonium salts is given in Supplementary Table 2. This Table 2 gives an overview of fragments/species belonging to the different ammonium salts. In the case of $NH_4OCN$, we use the same abundances for $NH_3$ as for $NH_4Cl$, for the fragments of HNCO we use data from NIST. We don't know how much $NH_4OCN$ is expected as the data in Hand et al. (42) are mostly qualitative. For $NH_4CN$ we use the same fragmentation pattern as for $NH_4Cl$ with unknown amounts of $NH_2CN$ and assuming that all CN is due to fragmentation of HCN (39).

Acetic acid $C_2H_4O_2$, the main species resulting from sublimation of ammonium acetate, has a stable isomer in the form of glycol aldehyde. However, they can be distinguished by their fragmentation pattern. Their fragmentation patterns for electron impact are quite different. For glycol aldehyde the most prominent fragment is $m/z$ 31 ($CH_3O$, 1400%), followed by $m/z$ 32 ($CH_4O$, 682%) (39), the numbers giving the relative abundance compared to the parent ion on $m/z$ 60. The major species expected from ammonium acetate is acetic acid with the major fragments $CH_3CO$ ($m/z$ 43, 153%) and $CHO_2$ ($m/z$ 45, 137%) (39).

Amines and other hydrogenated species

For amines of the form $H_nCN$ and $CH_mCN$, the abundance of fragments is high compared to the amount of saturated species ($H_5CN$ and $CH_7CN$) and does not correspond to fragmentation patterns from electron impact ionization of methylamine $CH_5N$ or ethylamine

$C_2H_7N$, respectively (Extended Data Figure 4). This points to hydrogenation of HCN or $CH_3CN$ rather than fragmentation of the saturated amines. Upon sublimation of ammonium salts hydrogen is released quite abundantly (e.g. $NH_3CO$ (formamide) => HNCO + 2H; $NH_4Cl$ => $NH_2Cl$ + 2H; etc.) and also during electron impact ionization ($NH_3 + e^- => NH_2^+ + H + 2e^-$). Similar patterns for the $CH_mN$ abundance distribution were found in the lab upon sublimation of pure ammonium formate (21) which took place outside of the mass spectrometer. Theulé et al. (43) showed that HCN hydrogenates quite easily when adding warm hydrogen atoms (300 K) to cold (15 K) HCN ice. They monitored *m/z* 27 to 31 while warming up the ice and found clear evidence for $CH_5N$. However, it is not completely clear from their data if hydrogenation really happened at 15 K or if hydrogen was trapped in the cold ice and then reacted only at warm-up of the sample.

Another indication of ongoing hydrogenation / protonation in our measurements is the high abundance of protonated $CH_4O$ and hydrogen sulfide.

Overall densities and dust impact
Starting at 17:50h on Sept. 5 the star trackers of the spacecraft reported high dust signals. The total density measured by ROSINA COPS showed noise peaks, which are typical for impacting dust and the overall density was much higher than in previous ellipses. This points to quite strong dust impacts on the spacecraft. At least one relatively big dust grain entered the DFMS ionization region at 18:14 h, blocking the ionizing electron current for a short time. The instrument, at that time was taking spectra in high resolution from m/z 13 to m/z 180 with 20s integration time per integer mass. The impact happened between the measurements of m/z 32 and m/z 33. The ion source at that time had a temperature of 273K.

Extended Data Figure 2, upper panel shows the total gas density at the spacecraft position as a function of time measured by ROSINA-COPS for four of the ellipses towards end of the mission. Shortly after 17h UTC on Sept. 5, 2016 (day 766) the pattern changes. Extended Data Figure 2, lower part shows the total density on Sept. 5 (day 766) and on Aug. 30 (day 760). The curve on Aug. 30 is shifted by 40 minutes as the closest approach was not exactly at the same time for the two ellipses. The sudden increase of the filament current regulating the electron emission in the ionization box to the nominal 200 μA (Extended Data Figure 2, blue curve) shows the dust impact.

As seen by ROSINA-DFMS the enhanced $NH_3$ density measured after the dust impact decreased steadily over the next few hours (Extended Data Figure 3). 14h after the impact, it had decreased by a factor 10 relative to the value at 20:17h. Due to a reaction wheel offloading of the S/C, DFMS had to be switched off at that time. When it was operational again, ~18h after the impact, the excess $NH_3$ had disappeared. This means that the ammonium salts had mostly sublimated after this time and at the ion source temperature of 273K.

At the time of the impact the S/C was 1.9 km above the cometary surface and at 3.7 AU from the Sun. Dust grains travel with a velocity of 1-10 m/s, which means they needed less than an hour to reach ROSINA, while most of the time during the mission, distances were well above 10 km and grains needed several hours to days to reach the spacecraft from the nucleus much closer to the Sun. It is therefore evident that during most of the mission time grains reaching the S/C had already lost their volatiles. COSIMA, the Cometary Secondary Ion Mass Analyzer (44), which was capturing dust grains and analyzing them by Secondary Ion Mass Spectrometry (SIMS), never detected ammonium salts.

COSIMA was collecting dust grains during extended periods ranging from several days to weeks on metal targets. The second step was then the imaging of the targets by COSISCOPE, a microscope. After download of the images, it was manually decided which grains to analyze by SIMS, the commands were uploaded and executed at the next opportunity. This means that between collecting grains and analysis, the grains stayed at ca. 283 K for at least several days, but mostly weeks or even months. This is another reason that COSIMA, complementary to ROSINA, did not detect (semi)-volatile molecules, but exclusively refractory macromolecules and minerals (45).

Data availability

The datasets analysed during the current study together with a user manual for data analysis are available in the ESA-PSA archive (https://www.cosmos.esa.int/web/psa/rosetta) or the NASA PDS archive (https://pdssbn.astro.umd.edu/data_sb/missions/rosetta/index.shtml)

**Additional references Methods**

**Extended Data**

Extended data figure 1: Sample DFMS spectra for m/z 60

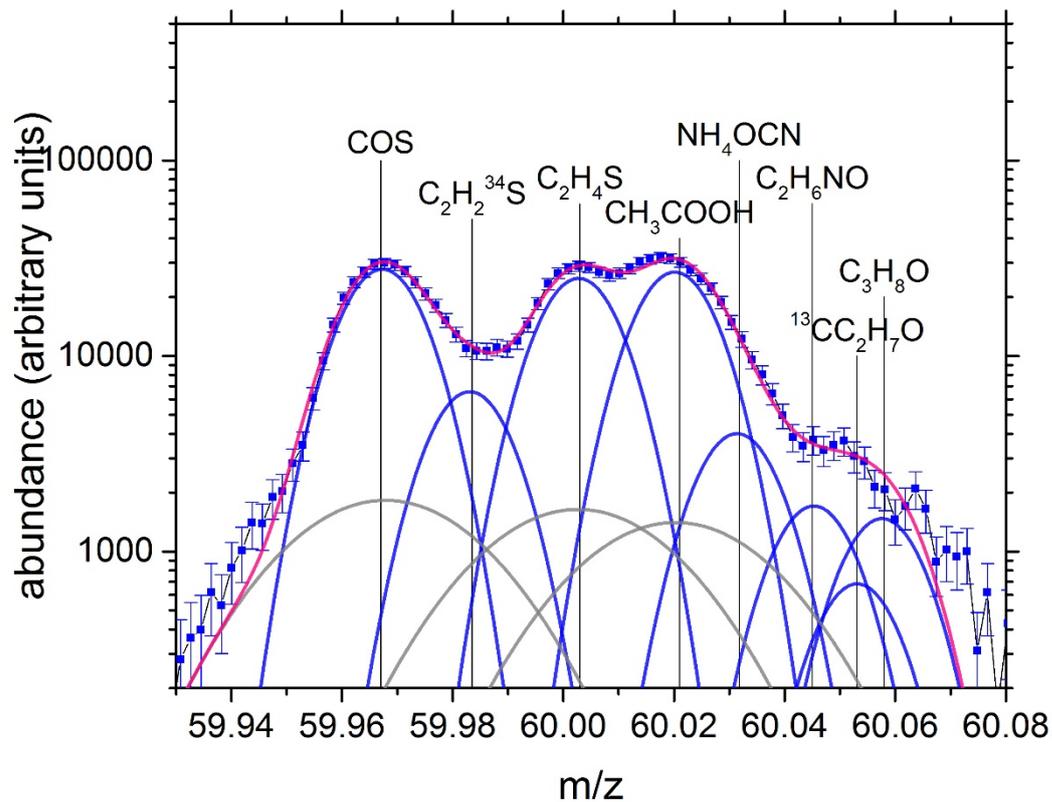

ROSINA-DFMS mass spectrum, Sept. 5, 2016, 18:34h. Error bars are 1-σ statistical errors. Blue/Grey curves are the two Gaussians, which describe the peaks, sharing the width across the spectrum.

Extended data figure 2: Total densities during the end of mission ellipses

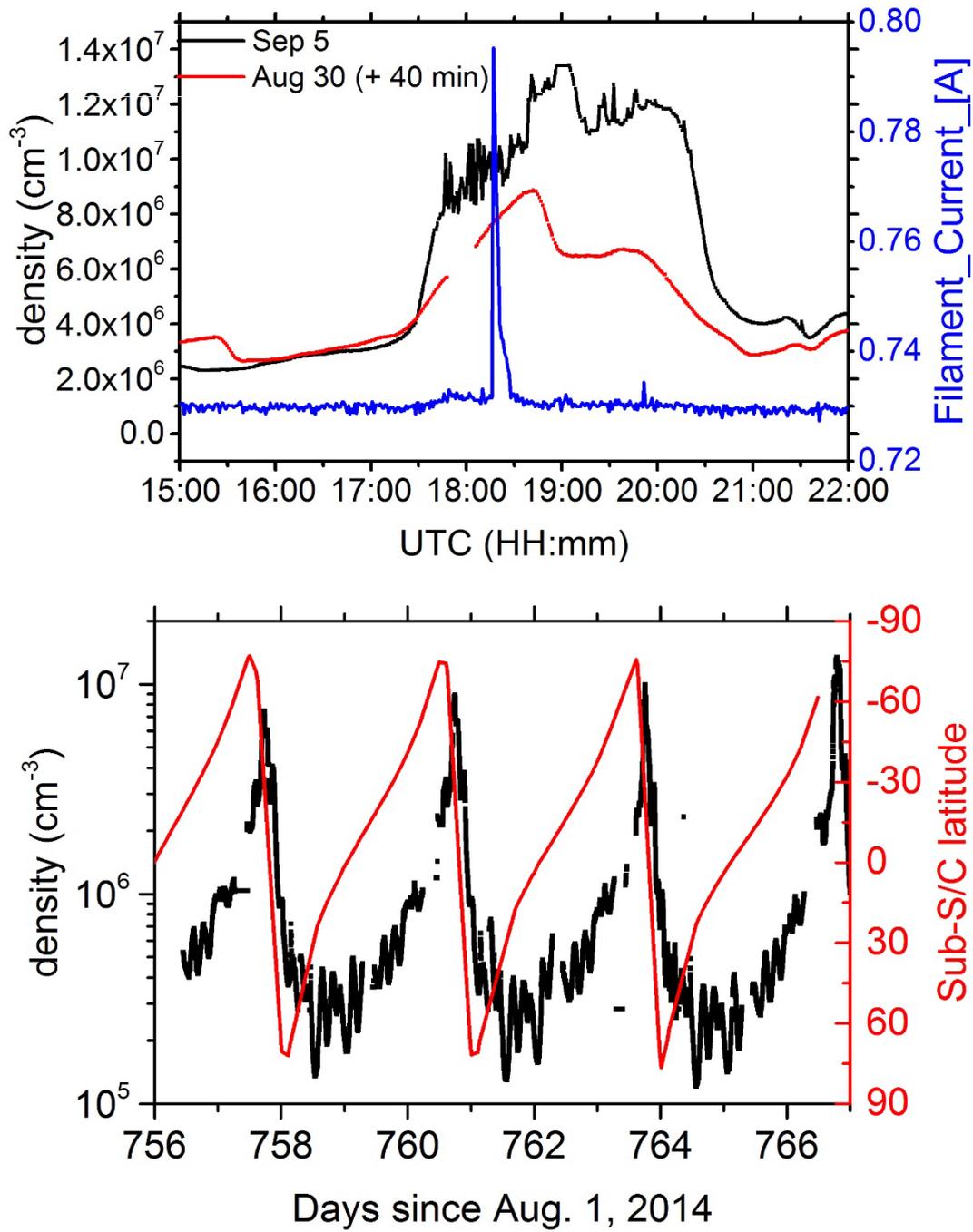

Total density from Aug. 26 to Sept. 5, 2016 (upper panel) and a zoom for Sept. 5, 15 h – 22 h UTC, measured by ROSINA-COPS. Also displayed are sub-spacecraft latitude in red and the filament current of DFMS for Sept. 5 in blue.

Extended data figure 3: Ammonia density with time on 5./6. Sept. 2016

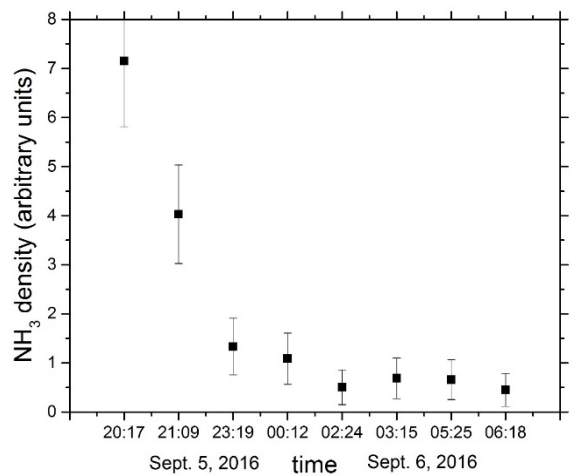

Ammonia density as a function of time on 5./6. September 2016.

Extended data figure 4: Amines and their fragments

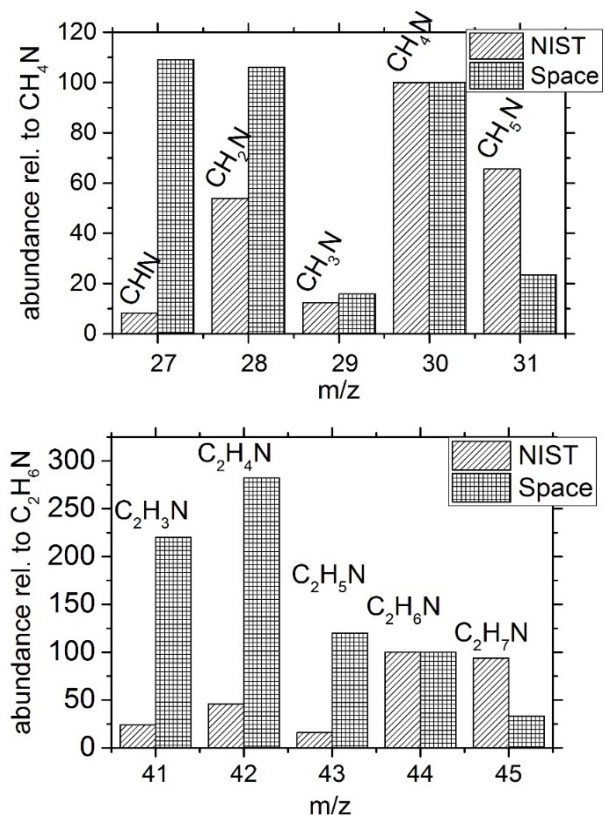

Comparison of methylamine and ethylamine fragments from electron impact ionization according to NIST and from measurements in space (Sept. 5, 2016, 20:19h).